\documentclass[preprint2]{aastex}

\def\agt{\mathrel{\raise.3ex\hbox{$>$}\mkern-14mu\lower0.6ex\hbox{$\sim$}}}
\def\alt{\mathrel{\raise.3ex\hbox{$<$}\mkern-14mu\lower0.6ex\hbox{$\sim$}}}

\newcommand{\beq}{\begin{equation}}
\newcommand{\eeq}{\end{equation}}
\newcommand{\beqn}{\begin{eqnarray}}
\newcommand{\eeqn}{\end{eqnarray}}

\shorttitle{Collapse of Rotating Stars}
\shortauthors{Shapiro}

\begin{document}

\title{Collapse of Uniformly Rotating Stars to 
Black Holes and the Formation of Disks}

\author{Stuart L. Shapiro \altaffilmark{1,2}}

\affil
{\altaffilmark{1} 
Department of Physics, University of Illinois at Urbana-Champaign,
\break
Urbana, IL 61801-3080}
\affil
{\altaffilmark{2} 
Department of Astronomy and NCSA, University of Illinois at
Urbana-Champaign,
\break
Urbana, IL 61801-3080}

\begin{abstract}
Simulations in general relativity show that the outcome of collapse
of a marginally unstable, uniformly rotating star spinning 
at the mass-shedding limit depends
critically on the equation of state. For a very stiff equation of state, 
which is likely to characterize a neutron star, essentially all of the 
mass and angular momentum 
of the progenitor are swallowed by the Kerr black hole 
formed during the collapse, leaving
nearly no residual gas to form a disk. For a soft
equation of state with an adiabatic index $\Gamma - 4/3 \ll 1$, which 
characterizes a very massive or supermassive star supported predominantly by 
thermal radiation pressure, as much as
$10\%$ of the mass of the progenitor avoids capture and 
goes into a disk about the central hole. We present a 
semi-analytic calculation that corroborates these
numerical findings and shows how the final outcome of such a collapse 
may be determined from simple physical considerations. 
In particular, we employ a simple energy variational principle 
with an approximate, post-Newtonian 
energy functional to determine the structure of
a uniformly rotating, polytropic star at the onset of collapse 
as a function of polytropic index $n$, where $\Gamma = 1+1/n$. 
We then use this data to
calculate the mass and spin of the final black hole and ambient disk.
We show that the fraction of the total mass that remains in the disk falls off sharply as 
$3-n$ (equivalently,  $\Gamma - 4/3$) increases.

\end{abstract}


\keywords{black hole physics -- relativity -- hydrodynamics --
stars: rotation}


\section{Introduction}

Determining the final state of a rotating star undergoing gravitational 
collapse is an important issue in relativistic astrophysics. Such a collapse
is the principle route by which a rotating black hole forms in nature.
Several different scenarios involving the collape of rotating stars to 
black holes have been the focus of considerable attention recently.
The collapse of massive stars in hypernovae explosions (``collapsars'')
may be the origin of long-period gamma-ray bursts (MacFadyen \& Woosley 1999; MacFadyen, Woosley
\& Heger 2001). The 
remnant of a binary neutron star, following inspiral and coalescence,
is likely to be rapidly rotating and 
undergo collapse, either prompt or delayed, depending on its mass (Shibata \& 
Uryu 2000, 2002; Baumgarte, Shapiro \& Shibata 2000). 
Binary neutron star merger and collapse provides a plausible
scenario for generating short-period gamma-ray bursts (Narayan, Paczynski \& Piran 1992; 
Ruffert \& Janka 1999) 
and a good candidate for the detection of gravitational
waves by Advanced LIGO and other high-frequency, gravitational wave laser 
interferometers.  The collapse of a rotating supermassive
star may be a promising route for forming the seeds of  
supermassive black holes (see Shapiro 2003 for a review and references), which are likely to reside at the centers of many, and
perhaps most, bulge galaxies (Richstone et al. 1998; Ho 1999), 
including the Milky Way 
(Genzel et al. 1997; Ghez et al. 2000; Sch\"odel et al. 2002) and are believed to be the engines that
power active galalactic nuclei (AGNs) and quasars (Rees 1998, 2001).

General relativity induces a radial instability in compact stars to catastrophic collapse.
Rotation can support larger masses in stable equilibrium, but all stars become unstable 
to collapse when they are sufficiently compact. Uniformly rotating stars can have masses exceeding the maximum allowed mass of 
a nonrotating spherical star by about $\lesssim 20\%$. 
Cook, Shapiro \& Teukolsky (1994a,b) have constructed numerical solutions 
of such ``supramassive'' stars, both for polytropes and for 14 realistic candidate nuclear equations of state.
Differentially rotating stars 
can be ``hypermassive'', with masses exceeding the maximum value for supramassive stars by factors or two or more 
(Baumgarte, Shapiro \& Shibata 2000; Lyford, Baumgarte \& Shapiro 2003). Viscosity (molecular or turbulent) and magnetic fields 
drive differentially rotating configurations to uniform rotation on secular 
timescales. 
[See, e.g., Shapiro 2000 and 
Cook, Shapiro \& Stephens 2003 for recent Newtonian calculations 
of magnetic braking and viscous damping of differential rotation,
and  Liu \& Shapiro 2004 for approximate, general relativistic
calculations of these phenomena; 
see Duez et al. 2004 
for detailed, general relativistic, numerical simulations of viscous damping of 
differential rotation.] 

Here we focus on uniformly rotating stars and consider the collapse of supramassive stars. 
The first investigation
of the gravitational collapse of supramassive stars  
was performed by Shibata, Baumgarte and Shapiro (2000), who 
employed a fully relativistic hydrodynamics code in three spatial dimensions plus time.
They studied the collapse of a marginally unstable, 
relativistic polytrope with polytropic index $n=1$ spinning at the mass-shedding limit. 
The mass--shedding limit is the maximal spin rate for a uniformly rotating star; here 
matter on the equator moves in
a Keplerian (geodesic) orbit about the star, supported against gravity 
entirely by centrifugal forces. Stars with $n=1$ 
have stiff equations of state, are not very centrally condensed and are fairly relativistic 
at the onset of collapse, with 
$R_e/M \approx 5.6$, where $R_e$ is the equatorial (circumferential) radius and $M$ 
is the total mass-energy.
Such stars provide crude models of rapidly rotating, massive
neutron stars. The outcome of collapse is a Kerr black hole containing all the rest-mass $M_0$ of the initial configuration,
and essentially all of the total mass-energy $M$ and angular momentum $J$ of the initial configuration
(apart from the small amount radiated away by gravitational waves). The reason that no matter remains
outside the hole is that at the equator the specific angular momentum of the gas, which is 
strictly conserved during axisymmetric collapse, is smaller than 
$j_{\rm ISCO}$, the
specific angular momentum of a particle at the innermost stable circular
orbit {\rm (ISCO)} about the final hole. Hence all the interior matter in the star 
is captured by the hole. 
This situation is similar to the collapse of a nonrotating spherical star, which forms a Schwarzschild black hole
without any ambient disk.  By contrast, Shibata \& Shapiro (2002) 
performed a fully relativistic hydrodynamics simulation in axisymmetry of the
collapse of a marginally unstable $n=3$ polytrope at the mass-shedding limit. Such a configuration  
models the likely endpoint state of an evolved, radiation-dominated, very massive or supermassive 
star at the onset of collapse (Baumgarte \& Shapiro 1999).  By contrast with an $n=1$ polytrope, 
an $n=3$ configuration
has a soft equation of state, is very centrally condensed and is 
nearly Newtonian ($R_e/M \approx 620$) 
at the onset of collapse.  A rotating black hole again forms during the collapse, 
but in this case the mass of the hole contains only about
$90\%$ of the total rest-mass of the system (with a 
spin parameter $J_h/M_h^2 \sim 0.75$). The remaining gas forms a
rotating disk about the nascent hole.
Here the angular momentum of the infalling matter in the outermost layers of the collapsing star 
exceeds $j_{\rm ISCO}$,  
so the gas in the outer regions escapes capture and forms a disk. This result has been corroborated by
a semi-analytic calculation by Shapiro \& Shibata (2002), which we 
generalize below for arbitrary $n$.

Recently, Shibata (2003) performed systematic numerical simulations in axisymmetry
to investigate the effects of the stiffness of the equation of state on the outcome of
the collapse of marginally unstable, supramassive stars spinning at the mass-shedding limit.
He studied collapsing relativistic polytropes with indicies $2/3 \leq n \leq 2$,
representing stiff and moderately stiff equations of state. He concluded that
for these configurations the final state is a Kerr black hole and the disk mass, if it exists at all, is 
very small ($ < 10^{-3}$ of the initial mass).

Simulations performed to date clearly show that the outcome of the collapse of a 
maximally spinning, supramassive star depends critically on the equation of state. In this paper 
we employ a straightforward semi-analytic analysis to corroborate this conclusion  
and to show how the final outcome of these detailed 
simulations may be inferred from 
simple physical considerations.
In particular, we employ an energy variational principle to determine the approximate structure of 
the marginally unstable, rotating star at the onset of collapse as a function of polytropic index $n$ and use this data to 
estimate the mass and spin of the final black hole and ambient disk. 
We show that the mass of the disk depends sensitively on the difference $3-n$.
Our semi-analytic calculations are based on an approximate post-Newtonian energy functional valid for rotational energies that
are much smaller than gravitational potential energies. They are most accurate for soft equations of state with $n$ close to 3, since the 
parameters of the progenitor star
are nearly Newtonian in this case and the effect of rotation in distorting the shape of the
interior region containing most of the mass is minimal. Interestingly, stars with $n$ close to 3
are the most difficult to simulate with a relativistic hydrodynamics code, since the large
size of the progenitor star requires that any such code span an enormous dynamic 
range and integrate for a huge number of (Courant) timesteps in order to follow the collapse to completion and resolve
the central black hole and disk reliably. In this sense, the two approaches -- the semi-analytic analysis
presented below and full numerical simulations -- are complementary.

\section{The Marginally Unstable, Supramassive Progenitor Star}
\label{Sec3}

In this Section we analyze the equilibrium and stability of a supramassive
polytrope spinning at mass-shedding and
identify the critical configuration at which radial
instability sets in. Our treatment is a straightforward generalization to arbitrary polytropic 
index $n$ of the semi-analytic analyses of Baumgarte \& Shapiro (1999).
We briefly review the predictions of the Roche
approximation for the rotating envelope in Section~\ref{roche}.  
In Section~\ref{crit} we
present a post-Newtonian energy variational calculation
which allows us to determine the critical configuration for each $n$ and identify its 
two key parameters, 
$R_p/M$ (where $R_p$ is the polar radius) and $J/M^2$, needed to determine the mass and spin of the 
final black hole and disk. We calculate the mass and spin of the black hole 
and disk in Section~\ref{disk}. Henceforth we adopt geometrized units and
set $G=c=1$.

\subsection{Review of the Newtonian Roche Model}
\label{roche}

Stars with soft equations of state are extremely centrally condensed:
they have an extended, low density envelope, while the bulk of the
mass is concentrated in the core.  For an $n=3$ Newtonian polytrope, for
example, the ratio between central density to average density is
$\rho_c/ \bar \rho = 54.2$.  The gravitational force in the envelope
is therefore dominated by the massive core, and it is thus legitimate
to neglect the self-gravity of the envelope.  In the equation of
hydrostatic equilibrium,
\begin{equation} \label{hydro}
\frac{{\bf \nabla} P}{\rho} = - {\bf \nabla} (\Phi + \Phi_c),
\end{equation}
this neglect amounts to approximating the Newtonian gravitational potential $\Phi$ by
\begin{equation} \label{potential}
\Phi = - \frac{M}{r}
\end{equation}
In~(\ref{hydro}) we introduce the centrifugal potential $\Phi_c$,
which, for constant angular velocity $\Omega$ about the $z$-axis, can
be written
\begin{equation} \label{centrif}
\Phi_c = - \frac{1}{2} \Omega^2\,(x^2 + y^2) = 
	- \frac{1}{2} \Omega^2 r^2 \sin^2 \theta.
\end{equation}
Integrating eq.~(\ref{hydro}) yields the Bernoulli integral
\begin{equation} \label{bernoulli}
h + \Phi + \Phi_c = H,
\end{equation}
where $H$ is a constant of integration and 
\begin{equation} \label{enthalpy}
h = \int \frac{dP}{\rho} = (n+1)\,\frac{P}{\rho}
\end{equation}
is the enthalpy per unit mass.  Here we have assumed a polytropic 
equation of state
\begin{equation}  \label{pressure}
P= K \rho^{\Gamma},  \mbox{~~~}
K, ~~\Gamma = 1 + 1/n = {\rm const},
\end{equation}
Evaluating eq.~(\ref{bernoulli})
at the pole yields
\begin{equation} \label{pole}
H = - \frac{M}{R_p},
\end{equation}
since $h = 0$ on the surface of the star and $\Phi_c = 0$ along the
axis of rotation.  In the following we assume that the polar radius
$R_p$ of a rotating star is always the same as in the nonrotating case. 
This assumption has been shown numerically to be very accurate (e.g.,
Papaloizou \& Whelan 1973). The mass of the star is hardly changed from its
value in spherical equilibrium, which to leading Newtonian order is given
by the polytropic relation
\beqn 
M & = &  4 \pi R_p^{(3-n)/(1-n)} \left[ \frac{(n+1)K}{4 \pi} \right]^{n/(n-1)} \nonumber \\
& & \times \ \xi_1^{(3-n)/(n-1)} \xi_1^2 |\theta'({\xi_1})|, \label{mass}
\eeqn
where the Lane-Emden functions for polytropes appearing above are tabulated in
Chandrasekhar (1939).
A rotating star reaches mass shedding when the equator orbits with the
Kepler frequency.  Using eqs.~(\ref{centrif}) and ~(\ref{bernoulli}), it 
is easy to show
that at this point the ratio between equatorial and polar radius is
\begin{equation}
\left( \frac{R_e}{R_p} \right)_{\rm shedd} = \frac{3}{2}.
\end{equation}
The corresponding maximum angular velocity is
\begin{equation} \label{shedd}
\Omega_{\rm shedd} = \left( \frac{2}{3} \right)^{3/2}
	\left( \frac{M}{R_p^3} \right)^{1/2}
\end{equation}
(Zel'dovich \& Novikov 1971; Shapiro \& Teukolsky 1983).

Since the bulk of the matter is
concentrated in the core and hardly affected by the rotation,
the moment of inertia of inertia of the star barely changes with
rotation and is well approximated by the nonrotating
value
\begin{equation} \label{I}
I = \frac{2}{5} \,\kappa_n \,M R_p^2 ~,
\end{equation}
where the nondimensional coefficient $\kappa_n$ is tabulated for many different
polytropes $n$  
in Lai, Rasio and Shapiro (1993) and for $n=3$ and $n=2.5$ in Table 1.
The ratio
between the kinetic and potential energy at mass-shedding then becomes
\begin{eqnarray} \label{t_ms}
\left( \frac{T}{|W|} \right)_{\rm shedd} & = &
	\frac{(1/2)I\,\Omega_{\rm shedd}^2}
        {\left( 3/(5-n) \right)M^2/R_p} \nonumber \\
	& = & \left( \frac{8}{81} \right) \left( 1-n/5 \right)\, \kappa_n. 
\end{eqnarray}
This result predicts that $T/|W|$ of a maximally rotating
polytrope of index $n$ is a
universal constant, independent of mass, radius, or angular velocity.

\begin{deluxetable}{ccccccc}
\tablewidth{0pc}
\tablecaption{Values of the polytropic structure coefficients}
\tablehead{$n$ & $\kappa_n$ & $k_1$ & $k_2$ & $k_3$ & $k_4$ & $k_5$ } 
\startdata 
3.0  & 0.18839\tablenotemark{a} & 1.7558\tablenotemark{a}	& 0.63899\tablenotemark{a}
	&  1.2041\tablenotemark{a} & 0.91829\tablenotemark{b} &
	0.33121\tablenotemark{c} \\
2.5  & 0.27951\tablenotemark{a} & 1.4295\tablenotemark{a}	& 0.67623\tablenotemark{a}
	&  1.4202\tablenotemark{a} & 0.86334\tablenotemark{c} &
	0.29752\tablenotemark{c} \\


\tablenotetext{a} {Lai, Rasio \& Shapiro (1993)}

\tablenotetext{b} {Shapiro \& Teukolsky (1983)}

\tablenotetext{c} {J. C. Lombardi Jr. (1997), priv. comm.}

\enddata
\end{deluxetable}

Inserting 
eqs.~(\ref{potential}), 
(\ref{centrif}), and
(\ref{enthalpy})--(\ref{shedd}) 
into eq.~(\ref{bernoulli}) yields the density
throughout the extended envelope,
\beqn
\rho & = & \frac{\xi_1^{3-n} (\xi_1^2 |\theta'({\xi_1})|)^{n-1}}{4 \pi} 
\frac{M}{R_p^3} \left(
\frac{R_p}{r}-1 \right. \nonumber \\ 
& & {}+\left. \frac{4}{27}\frac{r^2}{R_p^2}{\rm sin^2 \theta}
\right)^n.
\label {rho}
\eeqn
The stellar surface is the boundary along which $\rho = 0$, and is thus defined
by the curve $r(\theta)$ given by
\begin{equation}
\frac{4}{27}\frac{r^3}{R_p^3}{\rm sin^2 \,\theta} - \frac{r}{R_p} +1 = 0.
\end{equation}
The solution to this cubic equation is given by
\begin{equation}
\frac {r(\theta)}{R_p}= \frac{3\,{\rm sin \,(\theta/3)}}{{\rm sin \,\theta}}.
\end{equation}

\subsection{Critical Configuration}
\label{crit}
To determine the equilibrium and stability of a rotating polytrope, we express
its total energy as the sum of the (Newtonian) internal
energy $U$, the potential energy $W$, the rotational energy $T$, a
post-Newtonian correction $E_{PN}$ and a post-post-Newtonian
correction $E_{PPN}$. Writing the terms in that order yields the energy
functional 
\begin{eqnarray} 
E(\rho_c;M,J) & = & k_1 K M \rho_c^{1/n} - k_2 M^{5/3} \rho_c^{1/3} \nonumber  \\[1mm]
	& & {} + k_3 j^2 M^{7/3} \rho_c^{2/3} 
	- k_4 M^{7/3} \rho_c^{2/3} \nonumber \\  
        & & {} - k_5 M^3 \rho_c,  \label{energy}
\end{eqnarray}
where $\rho_c$ is the central density and where 
we have defined $j \equiv J/M^2$ and have neglected corrections
due to deviations from sphericity.  This neglect is justified, since
these corrections scale with $T/|W|$, which according to~(\ref{t_ms})
is always very small for centrally condensed configurations.  
Even though the value of the
post-post-Newtonian correction $E_{PPN}$ is very small, this term is
crucial for determining the critical, marginally stable configuration
for $n \equiv 3$,
as emphasized by Zel'dovich and Novikov (1971) and 
Baumgarte \& Shapiro (1999). The values of the
nondimensional coefficients $k_i$  
are determined by quadratures over Lane-Emden functions.
They are listed in
Table 1 for $n=3$ and $n=2.5$;
in our analysis below we evaluate them
for arbitrary $2.5 \leq n \leq 3.0$ by linearly interpolating between these 
two values of $n$.

Note that for any polytrope, $K^{n/2}$ has units of length.  We can
therefore introduce nondimensional coordinates by setting $K = 1$ (see
Cook, Shapiro \& Teukolsky 1992).  We will denote values
of nondimensional variables in these coordinates with a bar
(for example $\bar M$).  Values of these quantities for any other 
value of $K$ can be recovered easily by rescaling with an appropriate
power of $K^{n/2}$; for example $M = K^{n/2} \bar M$ and
$\rho = K^{-n} \bar \rho$.

Taking the first derivative of eq.~(\ref{energy}) with respect to the
central density, holding $M$ and $J$ constant, yields the condition for hydrostatic equilibrium:
\begin{eqnarray} \label{equilibrium}
0 = \frac{\partial \bar E}{\partial x} & = & 
	(3/n)k_1 \bar M x^{(3/n-1)} - k_2 \bar M^{5/3} \nonumber  \\
	& &{}  + 2 k_3 j^2 \bar M^{7/3} x - 2 k_4 \bar M^{7/3} x \nonumber \\ 
        & &{} - 3 k_5 \bar M^3 x^2,
\end{eqnarray}
where $x = \bar \rho_c^{1/3}$.  For stable equilibrium, the second
derivative of eq.~(\ref{energy}) has to be positive.  A root of the second
derivative therefore marks the onset of radial instability:
\beqn 
0  =  \frac{\partial^2 \bar E}{\partial x^2} & = &
	(3/n)(3/n-1) k_1 \bar M x^{(3/n-2)} \nonumber \\
	& & {} + 2 k_3 j^2 \bar M^{7/3} - 2 k_4 \bar M^{7/3} \nonumber \\ 
        & & {} - 6 k_5 \bar M^3 x. 
\label{stability}
\eeqn
To find the critical configuration at mass-shedding, 
Eqs.~(\ref{equilibrium}) and ~(\ref{stability}) must be solved 
simultaneously for $x$ and $\bar M$ subject to the constraint
\begin{equation} \label{t}
\frac{T}{|W|}_{\rm shedd} = \frac{k_3 j^2 \bar M^{7/3} x^2}{ k_2 \bar M^{5/3} x}
	= \frac{k_3 j^2 \bar M^{2/3} x}{ k_2 },
\end{equation}
where the left hand side is given by eq.~(\ref{t_ms}).

Key parameters characterizing the critical configurations are given in
columns  $2-4$ of Table 2. The first row for each $n$ gives the
solution found by 
solving eqs.~(\ref{equilibrium}) and ~(\ref{stability}) simultaneously,
substituting eq.~(\ref{t}). The other rows for selected $n$ give the
results of a careful integration of the full Einstein equations of
general relativity for the critical equilibrium configuration. 
The key global parameters needed to calculate the final black hole mass fraction and
spin are the stellar compaction $R_p/M$ and spin $J/M^2$. The value of $R_p/M$ determined 
by the simple variational method
described above are seen to be most reliable for nearly Newtonian configurations
with $n$ close to 3. Nevertheless, as $n$ decreases below 3 and the equation of state stiffens,
the trend toward higher compaction at the critical point is clearly evident.
The Newtonian Roche model for the outermost layers yields values for $J/M^2$ that
are consistent with the exact model solutions, albeit these are 
somewhat low, even for $n$ close to 3.
The discrepancy arises from using eq.~(\ref{I}) for the moment of inertia, which neglects
the increase in $I$ arising from oblateness in the outer layers, and is strictly valid only for
extreme central mass concentration\footnote{A  recalculation of the exact model by
Shibata (2004) for $n=3$ yields $J/M^2 = 0.91$,  which removes much of the discrepancy.}.

\begin{deluxetable}{cccccccc}
\tablecaption{Supramassive star collapse vs. polytropic index $n$}
\tablewidth{4.3in}
\tablehead{
\multicolumn{4}{c}{Critical Supramassive Star} & 
\colhead{} &
\multicolumn{3}{c}{Final Black Hole and Disk} \\
\colhead{n} & \colhead{${\rm \bar M}$} & \colhead{${\rm R_p/M}$} & 
\colhead{${\rm J/M^2}$} & 
\colhead{} &
\colhead{${\rm M_h/M}$} & \colhead{${\rm J_h/M_h^2}$} &
\colhead{${\rm M_d/M}$}}
\startdata
3.00  & 4.56         & 456        & 0.876       &  & 0.89        & 0.60        & 0.11   \\
      & 4.57${}^a$   & 427${}^a$  & 0.97${}^a$  &  & 0.87         & 0.71       & 0.13   \\
      & 4.57${}^a$   & 427${}^a$  & 0.97${}^a$  &  & 0.9${}^c$   & 0.75${}^c$  & 0.1${}^c$  \\
2.95  & 3.83         & 196        & 0.602       &  & 0.97        & 0.52        & 0.029  \\
2.90  & 3.30         & 119        & 0.491       &  & 0.99        & 0.45        & 0.011  \\
      & 3.30${}^b$   & 122${}^b$  & 0.567${}^b$ &  & 0.99        & 0.53        & 0.014  \\
2.80  & 2.54         & 60.9       & 0.382       &  & 1.00        & 0.37        & $2.5\times 10^{-3}$   \\
2.70  & 2.02         & 37.7       & 0.325       &  & 1.00        & 0.32        & $5.3\times 10^{-4}$   \\
2.60  & 1.65         & 25.5       & 0.287       &  & 1.00        & 0.29        & $9.7\times 10^{-5}$   \\
2.50  & 1.38         & 18.2       & 0.259       &  & 1.00        & 0.26        & $1.2\times 10^{-5}$   \\
      & 1.29${}^b$   & 28.9${}^b$ & 0.389${}^b$ &  & 1.00        & 0.39        & $2.7\times 10^{-4}$   \\
\enddata

\tablenotetext{a}{Baumgarte \& Shapiro (1999), Table 2}

\tablenotetext{b}{Cook, Shapiro \& Teukolsky (1994a), Table 2, converting $R_e$ to $R_p$
         using  $R_p/R_e \approx 2/3$.}

\tablenotetext{c}{Relativistic hydrodynamics simulation by
                 Shibata \& Shapiro (2002)}

\end{deluxetable}

\section{Calculating the Black Hole Mass and Spin}
\label{disk}

In this section we generalize to arbitrary polytropic index $n$ 
the iterative method of 
Shapiro \& Shibata (2002) to 
calculate the mass and spin of the black hole and any disk that might form
during the collapse of  supramassive star. (They formulated the method for
$n=3$).
Consider the implosion of matter from the envelope onto the central
black hole formed from the collapse of the centrally condensed interior 
region.  If the specific angular momentum of the imploding envelope matter 
is below $j_{\rm ISCO}$, the
specific angular momentum of a particle at the innermost stable circular
orbit {\rm (ISCO)} about the hole, the matter will be captured. If the angular momentum of the infalling matter exceeds $j_{\rm ISCO}$,  
the matter will escape capture and continue to orbit outside the hole, 
forming a disk. This capture criterion is well-supported by the numerical
simulations of rotating collapse by Shibata \& Shapiro (2002) for $n=3$ 
stars and by Duez et al. (2004) for $n=1$ stars. In both cases 
the capture of fluid in orbits with  $j > j_{\rm ISCO}$ was demonstrated 
to be negligible. 
This criterion suggests a simple iterative scheme
for calculating the final mass and spin of the hole and disk from the initial
stellar density and angular momentum profile. First, 
guess the mass and spin of the hole, $M_h$ and $J_h$. For 
our initial guess, we shall take a black hole that has consumed
all the mass and angular momentum of the star, so that
$M_h/M = 1 $ and $J_h/M_h^2 = (J/M^2)_{\rm crit}$. Next, 
use the initial stellar density and angular momentum profiles
to ``correct'' this guess by calculating the escaping  mass and angular momentum
of the outermost envelope with specific angular momentum
exceeding $j_{\rm ISCO}$.  We note that
the value $j_{\rm ISCO}$ depends on $M_h$ and $J_h$. We then ``correct'' the 
black hole mass and spin by deducting the 
values of the escaping mass and angular momentum of the envelope material  
from the guessed values of  $M_h$ and $J_h$. We recompute $j_{\rm ISCO}$ for the
``corrected'' black hole mass and spin, and repeat the calculation 
of the escaping envelope mass and angular momentum until convergence is 
achieved. The calculation described exploits the theorem that for
an axisymmetric dynamical system, the specific angular momentum spectrum,
i.e., the integrated baryon rest-mass of all
fluid elements with specific angular momentum $j$ less than a given value
(e.g., $j_{\rm ISCO}$) is strictly conserved in the absence of viscosity
(Stark \& Piran 1987). Any viscosity, if present, is expected to be unimportant
on dynamical timescales, as required by the theorem.

For a Kerr black hole of mass $M_h$ and spin parameter $a=J_h/M_h$, the
value of $j_{\rm ISCO}$ is given by
\begin{equation} \label{Jisco}
j_{\rm ISCO} =\frac{\sqrt{M_h r_{\rm ms}} (r_{\rm ms}^2 - 2 a \sqrt{M_h 
r_{\rm ms}} + a^2) }{r_{\rm ms}(r_{\rm ms}^2 - 3 M_h r_{\rm ms}
+ 2 a \sqrt{M_h r_{\rm ms}})^{1/2} }~,
\end{equation}
where $r_{\rm ms}$ is the ISCO given by
\begin{equation}
r_{\rm ms} = M_h \{ 3 + Z_2 - [(3 - Z_1)(3 + Z_1 + 2Z_2)]^{1/2} \}, 
\end{equation}
where
\beqn
Z_1 & \equiv & 1 + \left 
(1 - \frac{a^2}{M_h^2} 
\right )^{1/3} 
\left [ 
\left (1 + \frac{a}{M_h} \right )^{1/3} + \right. \nonumber \\ 
& & {} \left.\left (1 - \frac{a}{M_h} \right )^{1/3}
\right ],
\eeqn
and
\begin{equation} \label{Z2}
Z_2 \equiv \left (3 \frac{a^2}{M_h^2} + Z^2_1
\right )^{1/2}
\end{equation}
(see, e.g., Shapiro \& Teukolsky 1983).
Clearly, the infalling gas corotates with the black hole.

The mass of the escaping matter in the envelope 
with $j > j_{\rm ISCO}$ is given by
\begin{equation} \label{mesc}
\Delta M = \int\!\!\int \, 2\pi \varpi d\varpi \rho ~,
\end{equation}
where the density is given by eq.~(\ref{rho}) and the 
quadrature is performed over all cylindrical shells in the star with
cylindrical radii $\varpi > \varpi_{\rm ISCO} = (j_{\rm ISCO}/\Omega)^{1/2}$.
Here we set $\Omega = \Omega_{\rm shedd}$ 
given by eq.~(\ref{shedd}). (The quantity computed in eq.~(\ref{mesc}) is
actually the escaping rest-mass, but when the envelope is nearly
Newtonian, as we assume here, we can neglect the small difference between
rest-mass and total mass-energy.) 
Defining $\bar{\varpi}=\varpi/R_p$ and $\bar{z}=z/R_p$
gives the nondimensional integral
\beqn \label{dM}
& & \Delta M/M  =  \xi_1^{(3-n)}\left( \xi_1^2 |\theta'({\xi_1})| \right)^{(n-1)} \\
& & \times \, \int\!\!\int \, {\bar{\varpi}} d {\bar{\varpi}} d \bar{z} \,
\left [\frac{1}{(\bar{\varpi}^2 + \bar{z}^2)^{1/2}} -1 +
\frac{4}{27}\bar{\varpi}^2
\right ]^n ~. \nonumber
\eeqn
Similarly, the angular momentum carried off by the escaping 
matter in the envelope is given by
\begin{equation}
\Delta J = \int\!\!\int \, 2\pi \varpi d\varpi \rho \,\Omega \varpi^2 \,
\end{equation}
or
\beqn \label{dJ}
& & \Delta J/M^2  =  
\xi_1^{(3-n)}\left( \xi_1^2 |\theta'({\xi_1})| \right)^{(n-1)}
\left( \frac{2}{3} \right )^{3/2}
\left( \frac{R_p}{M} \right )^{1/2} \nonumber \\
& & \times \, \int\!\!\int \, {\bar{\varpi}}^3 d {\bar{\varpi}} d \bar{z} \,
\left [\frac{1}{(\bar{\varpi}^2 + \bar{z}^2)^{1/2}} -1 +
\frac{4}{27}\bar{\varpi}^2 
\right ]^n,
\eeqn
where $R_p/M$ is given in Table 2 and where once again the integral
is performed over all cylindrical shells in the star with
$\varpi > \varpi_{\rm ISCO} = (j_{\rm ISCO}/\Omega)^{1/2}$.
Eqns.~(\ref{dM}) and ~(\ref{dJ}) agree with Shapiro \& Shibata (2002),
eqns.~(22) and ~(24), for $n=3$.

The mass and angular momentum of the black hole can then be determined
from eqs.~(\ref{dM}) and ~(\ref{dJ}) according to
\begin{equation} \label{Mh}
M_{h}/M = 1 - \Delta M/M
\end{equation}
and
\begin{equation} \label{Jh}
J_h/M_h^2 = \frac {\left ( J/M^2 - \Delta J/M^2 \right)}
{\left ( 1 - \Delta M/M \right )^2}.
\end{equation}
Once the iteration of 
eqs.~(\ref{Jisco}) -- ~(\ref{Z2}) with
eqs.~(\ref{dM}), ~(\ref{Mh}) and ~(\ref{Jh}) 
converges, the mass
of the ambient disk can be found from
\begin{equation}
M_{\rm disk}/M = \Delta M/M.
\end{equation}
Typically, convergence to better than 1\% is achieved after only four 
iterations.

\begin{figure}[t!]
\plotone{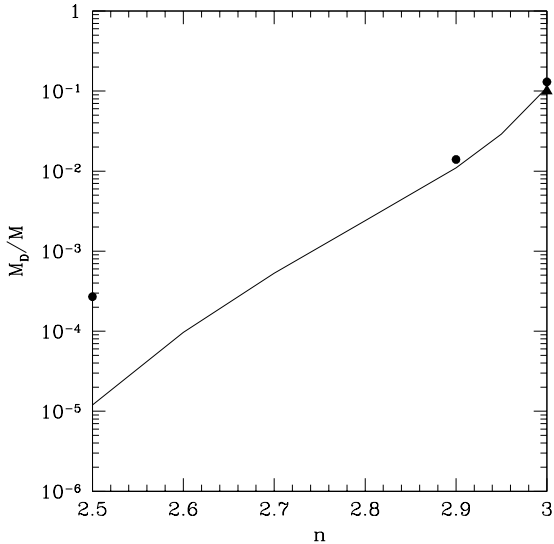}
\caption{Mass-fraction of the disk formed following collapse of a marginally unstable, 
supramassive polytrope at the mass-shedding limit versus polytropic index.
The solid line uses critical star parameters given by the variational treatment.
The solid dots use critical parameters determined from  numerical models of marginally 
unstable, relativistic equilibrium stars at the mass-shedding limit (see Table 2). 
The solid triangle shows the value determined by
a numerical simulation by Shibata \& Shapiro (2002).
\label{fig1}}
\end{figure}

\begin{figure}[t!]
\plotone{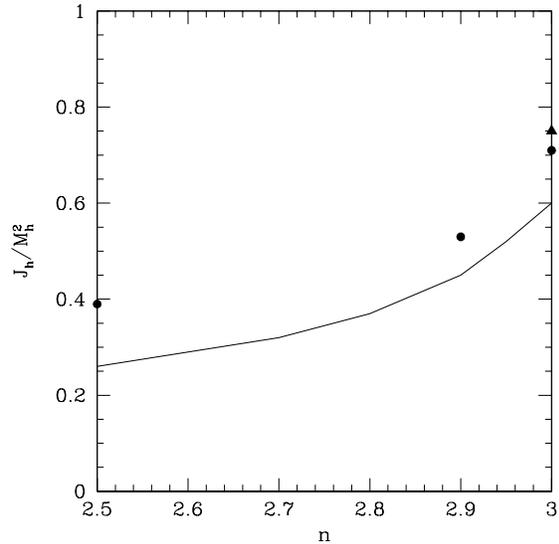}
\caption{Spin-parameter of the black hole formed following collapse of a marginally unstable, 
supramassive polytrope at the mass-shedding limit versus polytropic index. The curves and 
points are labelled as in Fig 1.
\label{fig2}}
\end{figure}

The calculated parameters for the final black hole and disk 
are given in the last three columns of Table 2 for different
values of $n$. The values for
$n=3$ are the most reliable, since the calculation of the
density and angular momentum distributions in the outermost layers
by means of a Newtonian Roche model is most accurate for 
such a soft equation of state. For $n=3$ we are able to compare the
semi-analytic calculations with the numerical collapse simulation of Shibata \&
Shapiro (2002).  Allowing for numerical error inherent in the 
simulation ($\lesssim$ few percent),
the agreement is quite good for the final masses of the black hole and disk, and 
reasonable, but less accurate, for the final black hole spin.
The numerical simulation (third row in Table 2)
used an initial model obtained by a numerical integration
of the full general relativistic, equilibrium equations
for the critical configuration at the onset of
collapse. When the parameters for this model are also used in the analytic
derivation of the final black hole and disk (second row in Table 2)
in place of the values determined by the variational principle (first row
in Table 2), the agreement with the spin is much improved.

The dependence of the black hole spin and disk mass on the polytropic index
is summarized in Figures 1 and 2. The spin parameter increases moderately with increasing
$n$ and increasing central concentration in the critical star. The disk mass is far more sensitive to the equation of state and increases very
rapidly with increasing $n$. There is hardly any mass in the disk unless $n$ is 
very close to $3$, a result which is consistent
with the dynamical simulations of Shibata, Baumgarte \& Shapiro (2000) and Shibata (2003) 
covering the range  $2/3 \leq n \leq 2$.

\section{Application: Very Massive Star Collapse}

A very massive, uniformly rotating star ($M/M_{\odot} \gg 100$) 
is characterized by the following properties:
\begin{enumerate}
\item It is dominated by thermal radiation pressure;
\item It is fully convective;
\item It is governed by nearly Newtonian gravitation;
\item It is described by the Roche model in the outer envelope.
\end{enumerate}

Such stars behave as polytropes with adiabatic
index $\Gamma = 1+1/n$ given by
\begin{equation}
\Gamma \approx 4/3 + \beta/6 + {\mathcal{O}} (\beta^2),
~~~~~~\beta \equiv \frac{P_{\rm g}}{P_{\rm r}},
\end{equation}
where
$P_{\rm r}= \frac{1}{3}a T^4$ is the radiation pressure and
$P_{\rm g}=nkT/\mu$ is the gas pressure (see, e.g., Shapiro \& Teukolsky 1983, 
Chapter 17). The ratio $\beta \ll 1$ is related to the  radiation entropy 
per baryon 
$s_{\rm r}$
($\approx$ constant throughout the star)
according to $\beta = (4/\mu)/(s_{\rm r}/k)$, where $\mu$ is the mean molecular weight,
while the mass is related to the
entropy according to 
\begin{equation}
s_{\rm r}/k \approx 0.942 (M/M_{\odot}).
\end{equation} 
Combining the above equations and evaluating $\mu \approx 2/(1+3X+0.5Y)$
for a zero-metallicity, primordial composition 
($X=0.75$ and $Y=0.25$;
Cyburt, Fields and Olive 2003) 
likely to characterize massive Population III objects, yields,  
to lowest order,
\begin{equation}
M/M_{\odot} \approx \frac{116}{(3-n)^2}.
\end{equation}

Baumgarte \& Shapiro (1999) have shown that, 
following cooling and contraction,
such a massive star will likely
settle into rigid rotation
and evolve to the mass--shedding limit, 
assuming the viscous or magnetic
braking timescale for angular momentum transfer
is shorter than the evolution timescale
in such an object 
(Bisnovatyi-Kogan et al. 1967; Zel'dovich \& Novikov 1971; Shapiro 2000; but see New \& 
Shapiro 2001 for an alternative scenario). 
Moreover,  Newtonian
simulations suggest that zero-metallicity rotating stars
with $M/M_{\odot} \gtrsim 260$ do not encounter the pair-production instability,
so they
collapse to black holes without exploding
(Fryer, Woosley \& Heger 2001).
The relation derived above between the mass $M$ and polytropic index $n$ 
for very massive stars 
then permits us to apply the results of the previous section to determine the
final fate of these objects after they reach the onset of radial instability 
and collapse.
For example, if the star has a mass $M/M_{\odot} \gtrsim 10^4$,
then $3-n \lesssim 0.1$. So according to Table 2 and Figs 1 and 2, the spin of the resulting hole will be 
moderate, $J_h/M_h^2 \gtrsim 0.5$, and the 
mass of the disk leftover from the collapse
will be nonnegligible, $M_D/M \gtrsim 10^{-2}$. The relativistic simulations of
Shibata \& Shapiro 2002 for pure $n=3$ polytropes
serve to confirm these predictions in the limiting regime
$3-n \ll 1$.

\section{Summary and Conclusions}

We have employed a simple analysis to determine the effect of the
stiffness of the equation of state on the fate of the collapse of a 
marginally unstable, relativistic polytrope spinning uniformly at the 
mass-shedding limit. We have used a variational principle and 
an approximate, post-Newtonian energy 
functional to determine the key parameters defining the 
structure of the critical progenitor star. We compared these parameters
with the results of detailed nummerical model calculations of stationary
configurations for select cases. We then employed a Roche model to
obain analytic expressions for the density and angular momentum profiles
in the envelope of the marginally stable star. We substituted these 
profiles into quadratures that determine the fractions of the 
stellar mass and angular momentum 
which escape capture by the central black hole assumed to 
form during the collapse.
The fraction of the progenitor mass and spin which go into the hole versus 
the ambient disk is then iterated until convergence is achieved.

We find that for the stars treated here, the mass fraction
in the disk is about $10\%$ for an $n=3$ polytrope and decreases rapidly
as $n$ decreases and the equation of state stiffens. The results are in 
agreement with the numerical simulations in $3+1$ by 
Shibata, Baumgarte \& Shapiro (2000) 
for $n=1$ and simulations by Shibata (2003) in axisymmetry 
for $2/3 \leq n \leq 2$, which show that the mass
fraction in a disk, if present at all, is less than $0.1\%$ of the total mass.
For the special case of $n=3$, the results are also in agreement with the 
simulation performed by Shibata \& Shapiro (2002) in axisymmetry, as has been
discussed previously by
Shapiro and Shibata (2002). The spin parameter of the black hole is less sensitive
to the stiffness of the equation of state, decreasing from $J_h/M_h^2 \approx 0.75$
for $n=3$ to $J_h/M_h^2 \approx 0.39$ for $n=2.5$.

The approach outlined here is more reliable for soft equations of state. In this
limit the marginally unstable configuration is nearly Newtonian and the post-Newtonian 
energy functional describing the bulk of the mass, the Newtonian Roche model describing
the envelope and the quadrature determining the escaping mass-energy fraction 
(assumed to equal the escaping rest-mass fraction) all become better approximations.
Interestingly, this is
precisely the limit where fully relativistic numerical simulations become more taxing, 
due to the
large dynamic range and very long time integrations required for a hydrodynamic
calculation.

While the semi-analytic approach presented here 
to predict the final black hole and disk masses and spins
was applied to treat the collapse of stars with {\it uniform} rotation, the method
can be used equally well to treat the collapse of unstable stars with {\it differential}
rotation. Recently, Duez et al. (2004) performed fully relativistic 
numerical simulations of 
hypermassive stars with appreciable differential rotation. 
Hypermassive stars
may form from the merger of binary neutron stars or from rotating core collapse
in supernovae (Baumgarte, Shapiro \& Shibata 2000). Magnetic braking or viscous damping of
differential rotation in such stars can drive them unstable to collapse on secular timescales.
Duez et al. (2004) performed simulations to demonstrate this effect in the case of viscosity 
and showed that the final black hole and disk
parameters are in agreement with the values predicted by the method outlined
in Section \ref{disk}. In particular, they showed that for rapid differential rotation characteristic
of hypermassive stars, the disk mass fraction is typically large ($10\% -- 20\%$ of the total
initial mass) for stiff equations of state with $n=1$. We therefore 
conclude that, in general, the parameters of the final black hole and the disk formed during the collapse of an unstable star 
depend both on the equation of state and on the degree of differential rotation.

\acknowledgments

It is a pleasure to thank T. W. Baumgarte, C. F. Gammie, and M. Shibata, for valuable
discussions.  This work was supported in part by NSF Grants PHY-0090310 and 
PHY-0205155 and NASA Grant NAG5-10781 at the
University of Illinois at Urbana-Champaign.

\end{document}